\documentclass{jltp}

\usepackage{graphicx} 

\title{Charge Transport Properties of Lightly-Doped Cuprates: Behavior
of the Hall Coefficient}

\author{Yoichi Ando, Kouji Segawa, A. N. Lavrov, and Seiki Komiya}

\address{Central Research Institute of Electric Power Industry,\\
Komae, Tokyo 201-8511, Japan}

\runninghead{Y. Ando {\it et al.}}{Charge Transport Properties of
Lightly-Doped Cuprates}

\begin{document}

\maketitle

\begin{abstract}

Behavior of $\rho_{ab}(T)$ and $R_H(T)$ is presented for LSCO and YBCO
single crystals in the lightly hole-doped antiferromagnetic region, with
an emphasis on the $R_H(T)$ data. In both systems, $R_H$ is virtually
constant at moderate temperatures and tends to increase at low
temperatures. Since essentially the same behavior of $\rho_{ab}(T)$ and
$R_H(T)$ is observed in both LSCO and YBCO, we discuss that the in-plane
charge transport properties are universal among the cuprates in the
lightly-doped regime and that the $R_H(T)$ data we obtained represent
the genuine behavior of the Hall effect in this regime.

\end{abstract}

\section{INTRODUCTION}

To elucidate the mechanism of high-$T_c$ superconductivity, it is
indispensable to understand how the metallic charge conduction emerges
and evolves from the parent Mott insulator as holes are doped to the
two-dimensional CuO$_2$ planes. Fueled by this motivation, we have been
intensively studying the lightly-doped region of the cuprates in the
past few years\cite{MR,Lavrov,mobility,sus,anisotropy,Nature,Komiya},
using high-quality single crystals of La$_{2-x}$Sr$_x$CuO$_4$ (LSCO) and
YBa$_2$Cu$_3$O$_y$ (YBCO). During the course of our research, it has
become evident that in the lightly-doped region, which is normally
called ``antiferromagnetic insulating regime", doped holes are
surprisingly mobile and show metallic transport at moderate temperatures
in sufficiently clean single crystals, even when the in-plane
resistivity significantly exceeds the Mott-Ioffe-Regel limit for
metallic 2D transport\cite{mobility}. Moreover, the in-plane motion of
the holes is found to be almost completely insensitive to the
establishment of the long-range antiferromagnetic (AF)
order\cite{MR,mobility}, although at low temperatures holes show strong
localization that can be characterized by a variable-range-hopping
behavior. Such anomalous metallic transport at moderate temperatures
suggests that charges are {\it mesoscopically} segregated from magnetic
domains to form a self-organized network of hole-rich
paths\cite{MR,mobility,sus,anisotropy}, which can be viewed as a nematic
or isotropic phase of fluctuating charge stripes\cite{Kivelson}.

It has been reported\cite{Noda} that the Hall resistivity of
La$_{1.4-x}$Nd$_{0.6}$Sr$_x$CuO$_4$ (LNSCO) tends to disappear upon
transition into the static stripe phase\cite{Tranquada}, and Noda {\it
et al.} has claimed\cite{Noda} that this is due to the intrinsically 1D
nature of the charge transport in the stripe phase. Thus, given that the
charge transport in the lightly-doped cuprates appears to be largely
governed by the stripes, it is natural to ask how the Hall coefficient
$R_H$ behaves in the lightly-doped region. Also, it was recently
reported by Wang and Ong\cite{Ong} that in a specially-treated samples
of YBCO in the AF regime, the Hall coefficient tends to vanish below the
N\'{e}el temperature $T_N$. Such observation was proposed to be due to
an intrinsic particle-hole symmetry in the system, which may well be
related to the charge stripes\cite{Emery,Prelovsek} or some other
peculiarities\cite{Onoda}. Therefore, it is important to scrutinize the
intrinsic behavior of $R_H$ in the lightly-doped AF regime of the
cuprates. In this paper, we show our data of $R_H(T)$ for lightly-doped
LSCO and YBCO, and discuss how we can sort out the genuine behavior of
the Hall coefficient in the lightly-doped cuprates.

\section{EXPERIMENTAL}

The high-quality LSCO single crystals are grown by the traveling-solvent
floating-zone (TSFZ) technique\cite{Komiya}. The LSCO crystals are
carefully annealed to remove excess oxygen, which is particularly
important for lightly-doped samples in ensuring that the hole doping is
exactly equal to $x$. The clean YBCO crystals are grown in Y$_2$O$_3$
crucibles by a conventional flux method\cite{Segawa}; to exclude the
conductivity contribution from the Cu-O chains, which run along the
$b$-axis in YBCO with $y \ge 6.35$, the crystals are detwinned\cite{60K}
at temperatures below 270$^{\circ}$C with an uniaxial pressure of
$\sim$0.1 GPa, and the resistivity is measured along the $a$-axis in
YBCO. The detwinning is done after the oxygen contents are tuned to the
desired values by proper annealing. Note that the detwinning is not
necessary for YBCO crystals with $y \le 6.30$, where the system has
tetragonal crystal symmetry. The in-plane resistivity $\rho_{ab}$ and
the Hall coefficient $R_H$ are measured using a standard ac six-probe
method. The Hall effect measurements are done by sweeping the magnetic
field to $\pm$14 T at fixed temperatures stabilized within $\sim$1 mK
accuracy. The Hall coefficients are always determined by fitting the
$H$-linear Hall voltage in the range of $\pm14$ T, which is obtained
after subtracting the magnetic-field-symmetrical magnetoresistance
component caused by small misalignment of the voltage contacts. Since we
always check for the linearity of the Hall voltage for each
magnetic-field sweep, we are confident that the Hall data shown here are
free from any adverse effect of superconducting fluctuations, which
could affect the results when samples contain a superconducting minor
phase.

\section{REMARKS ON YBCO SAMPLE PREPARATION}

It should be emphasized that the oxygen {\it arrangement} in the Cu-O
chain layers of YBCO is largely variable, which causes complications to
the study of YBCO. For example, for a given $y$ the actual hole doping
can differ depending on the arrangement of the O atoms, and the O atoms
in the Cu-O chains can rather easily rearrange at room temperature; this
causes the well-documented ``room-temperature (RT) annealing
effect"\cite{Lavrov}, with which the $T_c$ of heavily underdoped, but
still superconducting, sample slowly increases in several-day time
scale. Long-time annealing at 80 -- 150$^{\circ}$C can enhance
short-range oxygen ordering, but can never remove the intrinsic
randomness unless $\delta$ ($= 7-y$) is exactly 1/2 (for which
well-ordered ortho-II phase can be obtained). For this work (and in all
of our previous works\cite{MR,Lavrov,anisotropy,Segawa,60K,MR2}), the crystals
are always quenched at the end of the high-temperature annealing (which
tunes the oxygen content) and detwinning is performed at temperatures
below 220$^{\circ}$C after the annealing. The samples are left at room
temperature for at least a week for the local oxygen arrangement to
equilibrate; therefore, the oxygen atoms on the chain sites are locally
ordered (because of the RT annealing) but macroscopically uniform
(because of the quenching)\cite{Lavrov}. This procedure ensures very
good reproducibility of the transport properties, as has been
demonstrated in Ref. \onlinecite{60K}, and gives a sharp superconducting
transition\cite{60K} (for $y > 6.35$) or a sharp N\'{e}el
transition\cite{Lavrov} (for $y < 6.35$), both of which are hallmarks
of uniform hole distribution in the sample. Also, we have never observed
that a change in the annealing condition (including the RT annealing)
gives rise to a change in residual resistivity; this implies that the
randomness in the chain layers does not work as a strong disorder for
the in-plane charge transport, which is understandable because the Cu-O
chain layers are separated from the CuO$_2$ planes by the apical-oxygen
layers and thus are rather far ($\sim$4 \AA) away from the planes.

\section{RESULTS}

\begin{figure}
\centerline{\includegraphics[width=5in]{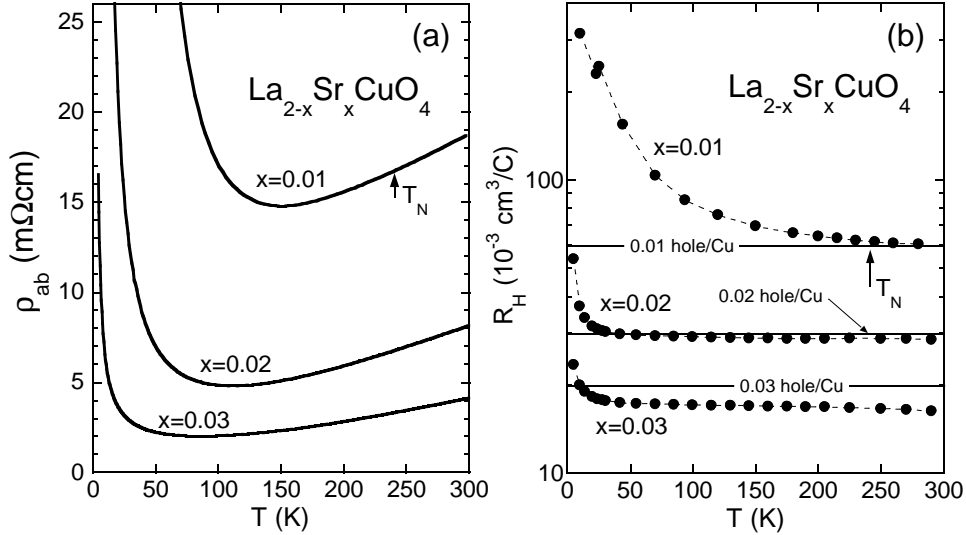}}
\caption{Temperature dependences of (a) $\rho_{ab}$ and (b) $R_H$ of 
lightly-doped LSCO single crystals ($x = 0.01 - 0.03$).
In panel (b), the values of $R_H^n$ expected for 
ordinary single-band metal with the three hole concentrations 
(0.01, 0.02, and 0.03 per Cu) are shown by solid lines.}  
\end{figure}

Figure 1 shows the in-plane resistivity $\rho_{ab}$ and the Hall
coefficient $R_H$ of lightly-doped LSCO single crystals ($x$ = 0.01 --
0.03). Note that the $x$ = 0.01 sample shows long-range N\'{e}el order
below 240 K (Ref. \onlinecite{mobility}). One can easily see in Fig.
1(a) that the $\rho_{ab}(T)$ data of all the samples show metallic
behavior ($d\rho_{ab}/dT > 0$) at moderate temperatures; in particular,
in the $x$ = 0.01 sample $\rho_{ab}(T)$ is completely insensitive to the
N\'{e}el ordering and keeps its metallic behavior well below $T_N$. This
insensitivity of the in-plane charge transport to N\'{e}el ordering may
not be so surprising, because the large $J$ ($\sim$0.1 eV) causes the
antiferromagnetic correlations to be well established in the CuO$_2$
planes from above $T_N$ (Ref. \onlinecite{Kastner}).

Figure 1(b) shows the corresponding $R_H(T)$ data for the three
concentrations. Note that $R_H$ is essentially temperature independent
in the temperature range where the metallic behavior of $\rho_{ab}(T)$
is observed, which is exactly the behavior that ordinary metals show.
Moreover, $R_H$ agrees well with the value $R_H^n$ expected for ordinary
single-band metal with the nominal hole concentration of $x$ per Cu
(i.e., $R_H^n \equiv V_0/ex$ where $V_0$ is the unit volume per Cu). At
low temperature where disorder causes the holes to localize, $R_H(T)$
reflects the localization (i.e., decrease in the number of {\it mobile}
holes) and tends to diverge, which is also an ordinary behavior for
disordered metals with low carrier density. Therefore, in the
lightly-doped LSCO, the behavior of $R_H$ appears to be a conventional
one for disordered metals; this is quite opposite to the behavior in the
superconducting regime, where $R_H$ significantly departs from $R_H^n$
and shows a strong temperature dependence.

\begin{figure}
\centerline{\includegraphics[width=90mm]{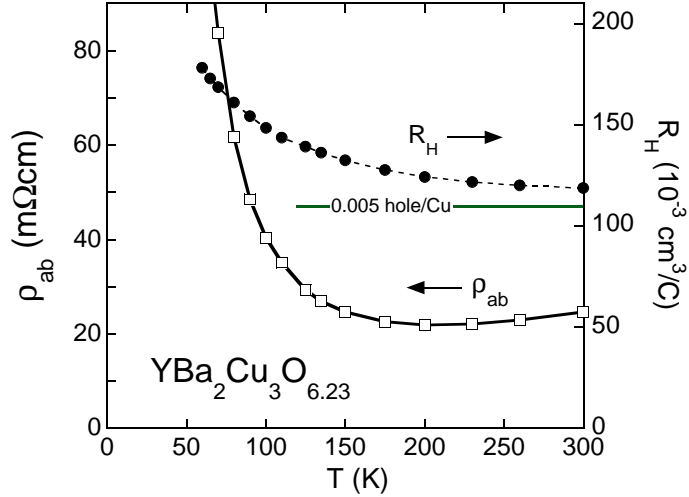}}
\caption{Temperature dependences of $\rho_{ab}$ (left axis) and 
$R_H$ (right axis) of slightly-doped YBCO at $y$ = 6.22.
The value of $R_H^n$ expected for ordinary 
single-band metal with hole concentration of 0.005 per Cu 
is shown by solid line.}  
\end{figure}

Figure 2 shows $\rho_{ab}(T)$ and $R_H(T)$ data of a very lightly
hole-doped YBCO single crystal ($y$ = 6.22); remember that in YBCO the
hole doping onto the CuO$_2$ planes starts above $y \simeq 6.20$. The
behavior of both quantities is essentially the same as that in
lightly-doped LSCO: $\rho_{ab}(T)$ shows a metallic behavior at moderate
temperatures and is insensitive to the N\'{e}el ordering (which occurs
above 300 K for this composition); $R_H$ is only weakly temperature
dependent in the temperature range where the metallic $\rho_{ab}(T)$ is
observed, and its magnitude is consistent with the expected $R_H^n$
value (we estimate the hole concentration per Cu, $p$, is less than 0.01
and probably around 0.005 for this composition). Also, $R_H(T)$ reflects
the charge localization and tends to increase at low temperature.
Therefore, our data indicate that the in-plane charge transport in the
lightly-doped region is essentially universal among LSCO and YBCO, and
that there is no hint of vanishing $R_H$ in the AF state.

\begin{figure}
\centerline{\includegraphics[width=5in]{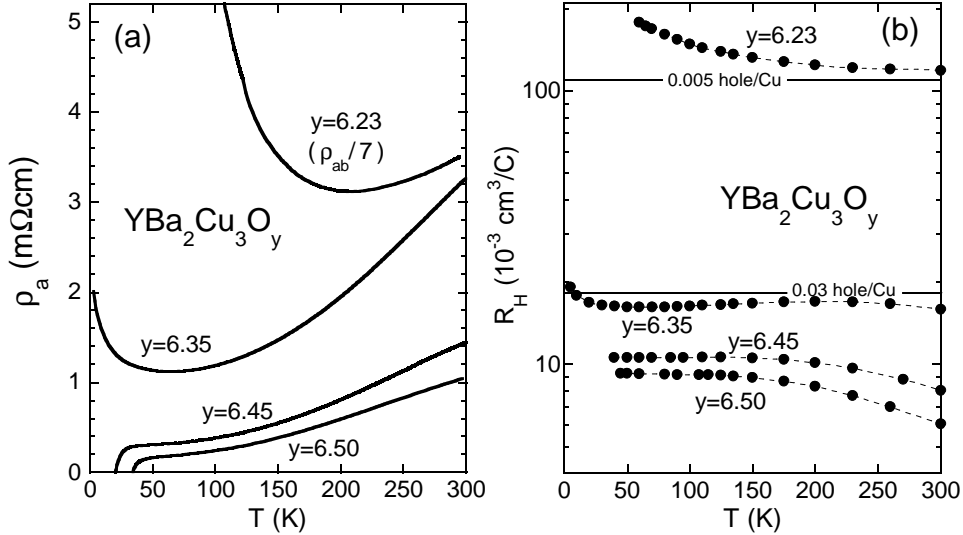}}
\caption{Temperature dependences of (a) $\rho_{a}$ and (b) $R_H$ of 
lightly-doped YBCO single crystals ($y$ = 6.22, 6.35, 6.45, and 6.50).
In panel (b), the values of $R_H^n$ expected for ordinary 
single-band metal with hole concentrations of 0.005 and 0.03 per Cu 
are shown by solid lines.}  
\end{figure}

For YBCO, we also show $\rho_{a}(T)$ and $R_H(T)$ data for higher, but
still heavily underdoped, concentrations (Fig. 3); note that the samples
for $y$ = 6.35, 6.45, and 6.50 are detwinned single crystals. We
estimate the $p$ value for $y$ = 6.35 to be around 0.03, and in Fig.
3(b) one can see that the measured $R_H$ value at moderate temperatures
are somewhat lower than, but close to, the expected $R_H^n$ value for
its hole concentration, just as the case of LSCO at $x$ = 0.03.
(Remember that in YBCO the actual $p$ value for a given $y$ varies
depending on the oxygen arrangement as already mentioned, and thus the
estimation of $p$ in YBCO can never be exact.)

\section{DISCUSSIONS}

\subsection{Hall Coefficient and Stripes}

As is mentioned in the Introduction, it has been shown that the Hall
effect tends to disappear in LNSCO in the static stripe
phase\cite{Noda}. Although such a result appears to naturally indicate
the 1D behavior at first sight, one should note that the quasi-1D motion
itself does {\it not} necessarily drive the Hall coefficient to zero.
The quasi-1D confinement indeed suppresses the transverse (Hall) {\it
current} induced by the magnetic field, but the same large transverse
resistivity restores the finite Hall {\it voltage}, because of the
relation $R_H \sim \sigma_{xy}/\sigma_{yy}\sigma_{xx}$. For the same
reason, for example, the strong charge confinement in the CuO$_2$ planes
in the cuprates does {\it not} prevent generation of the Hall voltage
along the $c$-axis when $H$ is applied along the $ab$
plane\cite{Harris}. The anomaly in the Hall effect in LNSCO is likely to
be caused by the peculiar arrangement of stripes which alter their
direction from one CuO$_2$ plane to another, thereby avoiding
$\sigma_{yy}$ to vanish. On the other hand, the unidirectional
stripes\cite{anisotropy,Matsuda} that exist in pure LSCO in the
lightly-doped region may keep the Hall coefficient unchanged. Also
possible is the finite Hall resistivity being caused by the transverse
(sliding) motion of the stripe as a whole; in fact, very recent optical
conductivity measurements of lightly-doped LSCO showed that the sliding
degrees of freedom are important for the realization of the metallic
transport in this system\cite{Dumm}. Therefore, even when the in-plane
charge transport is essentially governed by the stripes in the
lightly-doped cuprates\cite{mobility,anisotropy}, it is not at all
surprising that $R_H$ does not vanish and shows rather conventional
behavior.

\subsection{Antiferromagnetic YBCO}

As is shown in Fig. 3(b), $R_H(T)$ of our quenched YBCO samples does not
exhibit any pronounced peak, but rather shows an upturn with decreasing
temperature when the resistivity displays an insulating behavior; this
result is in clear disagreement with the result reported in Ref.
\onlinecite{Ong}, which was obtained for samples annealed at
150$^{\circ}$C for 3 weeks. In 1992, Jorgensen's group at Argonne
National Laboratory intensively studied\cite{Radaelli} the effect of
long-time annealing of YBCO; since it had already been
known\cite{Jorgensen} that in quenched samples the
tetragonal-to-orthorhombic transition occurs at $y \simeq 6.35$, they
carefully studied the samples which showed orthorhombicity down to $y =
6.25$ after a long-time annealing. They concluded that such
orthorhombicity was a result of a macroscopic phase separation into
tetragonal and orthorhombic phases; furthermore, they explicitly showed
that this macroscopic phase separation took place when samples with $y$
of around 6.3 were annealed at low temperature for a long time. The
mechanism of this phase separation can be understood by looking at the
phase diagram (such as that in Ref. \onlinecite{Zimmermann}); for
example, when a sample is at $y=6.30$ and is annealed at
150$^{\circ}$C for a long time, the proximity of the orthorhombic phase
(which is stabilized above $y \simeq 6.4$ at this temperature and
apparently has a lower free energy) in the phase diagram causes
oxygen-rich domains with the orthorhombic symmetry to be formed and the
rest of the sample is left to be oxygen-poor and tetragonal, leading to
a macroscopic phase separation. 

On the other hand, as the neutron scattering works have
shown\cite{Radaelli,Jorgensen}, the quenched samples have no such
problem as the macroscopic phase separation. In fact, the $R_H(T)$ data
in Fig. 2 are essentially consistent with the $R_H(T)$ behavior in
lightly-doped LSCO, which gives confidence that the $R_H(T)$ data we
obtained in the antiferromagnetic YBCO represent the genuine behavior of
the Hall coefficient of the lightly-doped cuprates. Since other in-plane
transport properties show striking similarities between LSCO and YBCO in
the lightly-doped region\cite{mobility,anisotropy}, it is actually more
natural to see the Hall effect to behave similarly in the two systems,
than to see quite different behaviors.

\section{SUMMARY}

High-quality data of $\rho_{ab}(T)$ and $R_H(T)$ are shown for LSCO and
YBCO single crystals in the lightly-doped region. The presentation is
focused on $R_H(T)$, which exhibits almost temperature-independent value
in the temperature range where $\rho_{ab}(T)$ shows a metallic behavior.
The value of $R_H$ in this nearly-temperature-independent regime agrees
rather well with the value expected from the hole concentration for
ordinary single-band metal. Notably, such behavior of $R_H(T)$ is
observed in both LSCO and YBCO in the lightly-doped region, suggesting
that the in-plane charge transport properties are universal among
cuprates in the AF regime; this gives us confidence that the $R_H(T)$
data we obtained represent the genuine behavior of the Hall effect in
the lightly-doped cuprates.

\section*{ACKNOWLEDGMENTS}

We thank N. P. Ong for collegial information exchange and discussions.
We also thank J. M. Tranquada for collaborations and discussions, and D.
A. Bonn, B. Keimer, S. A. Kivelson, N. Nagaosa, T. Tohyama, and S.
Uchida for helpful discussions.


\begin{thebibliography}{99}

\bibitem{MR}
Y. Ando, A. N. Lavrov, and K. Segawa, 
{\it Phys. Rev. Lett.} {\bf 83}, 2813 (1999).

\bibitem{Lavrov}
A. N. Lavrov, Y. Ando, K. Segawa, and J. Takeya, 
{\it Phys. Rev. Lett.} {\bf 83}, 1419 (1999).

\bibitem{mobility}
Y. Ando, A. N. Lavrov, S. Komiya, K. Segawa, and X. F. Sun, 
{\it Phys. Rev. Lett.} {\bf 87}, 017001 (2001).

\bibitem{sus}
A. N. Lavrov, Y. Ando, S. Komiya, I. Tsukada, 
{\it Phys. Rev. Lett.} {\bf 87}, 017007 (2001).

\bibitem{anisotropy}
Y. Ando, K. Segawa, S. Komiya, and A. N. Lavrov, 
{\it Phys. Rev. Lett.} {\bf 88}, 137005 (2002).

\bibitem{Nature}
A. N. Lavrov, S. Komiya, and Y. Ando, 
{\it Nature} {\bf 418}, 385 (2002).

\bibitem{Komiya}
S. Komiya, Y. Ando, X. F. Sun, and A. N. Lavrov, {\it Phys. Rev. B} 
{\bf 65}, 214535 (2002).

\bibitem{Kivelson}
S. A. Kivelson, E. Fradkin, and V. J. Emery, 
{\it Nature} {\bf 393}, 550 (1998).

\bibitem{Noda}
T. Noda, H. Eisaki, and S. Uchida, {\it Science} {\bf 286}, 265 (1999).

\bibitem{Tranquada}
J. M. Tranquada {\it et al.}, {\it Nature} {\bf 375}, 561 (1995).

\bibitem{Ong}
Y. Wang and N. P. Ong, {\it PNAS} {\bf 98}, 11091 (2001).

\bibitem{Emery}
V. J. Emery, E. Fradkin, S. A. Kivelson, and T. C. Lubensky,
{\it Phys. Rev. Lett.} {\bf 85}, 2160 (2000).

\bibitem{Prelovsek}
P. Prelovsek, T. Tohyama, and S. Maekawa, 
{\it Phys. Rev. B} {\bf 64}, 052512 (2001).

\bibitem{Onoda}
M. Onoda and N. Nagaosa, {\it Phys. Rev. B} {\bf 65}, 214502 (2002).

\bibitem{Segawa}
K. Segawa and Y. Ando, {\it Phys. Rev. B} {\bf 59}, R3948 (1999).

\bibitem{60K}
K. Segawa and Y. Ando, {\it Phys. Rev. Lett.} {\bf 86}, 4907 (2001).

\bibitem{MR2}
Y. Ando and K. Segawa, {\it Phys. Rev. Lett.} {\bf 88}, 167005 (2002).

\bibitem{Kastner}
M. A. Kastner, B. J. Birgeneau, G. Shirane, and Y. Endoh, 
{\it Rev. Mod. Phys.} {\bf 70}, 89 (1998).

\bibitem{Harris}
J. M. Harris, Y. F. Yan, and N. P. Ong, 
{\it Phys. Rev. B} {\bf 46}, 14293 (1992).

\bibitem{Matsuda}
M. Matsuda {\it et al.}, {\it Phys. Rev. B} {\bf 65}, 134515 (2002).

\bibitem{Dumm}
M. Dumm, D. N. Basov, S. Komiya, and Y. Ando, unpublished.

\bibitem{Radaelli}
P. G. Radaelli, C. U. Segre, D. G. Hinks, and J. D. Jorgensen,
{\it Phys. Rev. B} {\bf 45}, 4923 (1992).

\bibitem{Jorgensen}
J. D. Jorgensen {\it et al.}, {\it Phys. Rev. B} {\bf 41}, 1863 (1990).

\bibitem{Zimmermann}
M. v. Zimmermann {\it et al.}, cond-mat/9906251.

\end{thebibliography}
\end{document}